\documentclass{aa}
\usepackage{graphicx}
\usepackage{amsfonts,amssymb,latexsym}
\usepackage{epstopdf}

\def\lambdabar{\protect\@lambdabar}
\def\@lambdabar{%
\relax
\bgroup
\def\@tempa{\hbox{\raise.73\ht0
\hbox to0pt{\kern.25\wd0\vrule width.5\wd0
height.1pt depth.1pt\hss}\box0}}%
\mathchoice{\setbox0\hbox{$\displaystyle\lambda$}\@tempa}%
{\setbox0\hbox{$\textstyle\lambda$}\@tempa}%
{\setbox0\hbox{$\scriptstyle\lambda$}\@tempa}%
{\setbox0\hbox{$\scriptscriptstyle\lambda$}\@tempa}%
\egroup
}
\def\chem#1#2{$\rm{}^{#1}\kern-0.8pt#2$}
\def\reac#1#2#3#4#5#6{$\rm\,{}^{#1}\kern-0.8pt{#2}\,({#3}\,,{#4})\,
{}^{#5}\kern-0.8pt{#6}\,$}
\def\gsimeq{\,\,\raise0.14em\hbox{$>$}\kern-0.76em\lower0.28em\hbox
{$\sim$}\,\,}
\def\lsimeq{\,\,\raise0.14em\hbox{$<$}\kern-0.76em\lower0.28em\hbox
{$\sim$}\,\,}
\def\be{\begin{equation}}
\def\ee{\end{equation}}
\def\beqy{\begin{eqnarray}}
\def\eeqy{\end{eqnarray}}
\def\bmlet{\begin{mathletters}}
\def\emlet{\end{mathletters}}

\def\A&A#1#2#3{ {\it Astron. Astrophys.} {\bf #2}, #3 (#1)}

\begin{document}

% A&A Section 12: Atomic, molecular and nuclear data}

\title{Improved predictions of nuclear reaction rates with the TALYS reaction code for astrophysical applications}

   \author{S. Goriely \inst{1},  S. Hilaire \inst{2} \and A.J. Koning \inst{3}}

   \offprints{S. Goriely}

\institute{
Institut d'Astronomie et d'Astrophysique, Universit\'e Libre de Bruxelles,  Campus de la Plaine CP 226, B-1050 Brussels, Belgium
\and 
DPTA/Service de Physique Nucl\'eaire, CEA/DAM Ile de France, BP 12, 91680 Bruy\`eres-le-Ch\^atel, France 
\and 
Nuclear Research and Consultancy Group, P.O. Box 25, NL-1755 ZG Petten, The Netherlands
}

    \date{Received --; accepted --}

\abstract{Nuclear reaction rates of astrophysical applications are traditionally determined on the basis of Hauser-Feshbach reaction codes. These codes  adopt a number of approximations that have never been tested, such as a simplified width fluctuation correction, the neglect of delayed or multiple-particle emission during the electromagnetic decay cascade, or the absence of the pre-equilibrium contribution at increasing incident energies. }
{The reaction code TALYS  has been recently updated to estimate the Maxwellian-averaged reaction rates that are of astrophysical relevance. These new developments enable the reaction rates to be calculated with increased accuracy and reliability and the approximations of previous codes to be investigated. } 
{The TALYS predictions for the thermonuclear rates of relevance to astrophysics are detailed and compared with those derived by widely-used codes for the same nuclear ingredients. } 
{It is shown that TALYS predictions may differ significantly from those of previous codes, in particular for nuclei for which no or little nuclear data is available. The pre-equilibrium process is shown to influence the astrophysics rates of exotic neutron-rich nuclei significantly. For the first time, the Maxwellian-averaged (n,2n) reaction rate is calculated for all nuclei and its competition with the radiative capture rate is  discussed. }
{The TALYS code provides a new tool to estimate all nuclear reaction rates of relevance to astrophysics with improved accuracy and reliability.
}

\titlerunning{Nuclear reaction rates with the TALYS code}
\authorrunning{S. Goriely et al.}

\maketitle

\section{Introduction}
\label{sect_intro}
Strong, weak, and electromagnetic interaction processes play an essential
role in many different applications of nuclear physics, such as reactor physics, waste
incineration, production of radioisotopes for therapy and diagnostics, charged-particle
beam therapy, material analysis, and nuclear astrophysics. Although significant
effort has been devoted in the past decades to measuring reaction cross sections,
experimental data only covers a minute fraction of the entire data set required for
such nuclear physics applications. Reactions of interest often concern unstable or even
exotic (neutron-rich, neutron-deficient, superheavy) species for which no experimental
data exist. Given applications (in particular, nuclear astrophysics and accelerator-driven
systems) involve a large number (thousands) of unstable nuclei for which many different
properties have to be determined. Finally, the energy range for which
experimental data is available is restricted to the small range that can be studied by present
experimental setups. To fill the gaps, only theoretical predictions can be used. 

Reaction rates for medium or heavy nuclei, applied in astrophysical applications, have been calculated until now on the basis of the Hauser-Feshbach (HF) statistical model (Hauser \& Feshbach 1952), adopting simplified schemes to estimate the capture reaction cross section of a given target nucleus, not only in its ground state but also in the different thermally populated states of the stellar plasma at a given temperature (Arnould 1972; Holmes et al. 1976;  Woosley et al. 1978, Sargood 1982, Thielemann et al. 1987; Cowan et al. 1991; Rauscher \& Thielemann 2001; Aikawa et al. 2005).  Such schemes include a number of approximations that have never been tested, such as the neglect of delayed particle emission during the electromagnetic decay cascade or the absence of the pre-equilibrium contribution at increasing incident energies or for exotic neutron-rich nuclei. 

In parallel, the nuclear physics community has developed tools for specific applications, such as accelerator-driven systems, which can shed light on the many approximations in nuclear astrophysical applications. One of these tools is the modern reaction code called TALYS (Koning et al. 2002, 2004, 2007). TALYS is a software for the simulation of nuclear reactions, which includes many state-of-the-art nuclear models to cover all main reaction mechanisms encountered in light particle-induced nuclear reactions. TALYS provides a complete description of all reaction channels and observables and in particular  takes into account all types of direct, pre-equilibrium, and compound mechanisms to estimate the total reaction probability as well as the competition between the various open channels.  The code is optimized for incident projectile energies, ranging from 1~keV up to 200~MeV on target nuclei with mass numbers between 12 and 339. It includes photon, neutron, proton, deuteron, triton, $^3$He, and $\alpha$-particles as both projectiles and ejectiles, and single-particle  as well as  multi-particle emissions and fission.  All experimental information on nuclear masses, deformation, and low-lying states spectra is considered, whenever available (in particular the RIPL2 database; Belgya et al. 2006). If not, various local and global input models have been incorporated to represent the nuclear structure properties, optical potentials, level densities, $\gamma$-ray strengths, and fission properties. The TALYS code was designed to calculate total and partial cross sections, residual and isomer production cross sections, discrete and continuum $\gamma$-ray production cross sections, energy spectra, angular distributions, double-differential spectra,  as well as  recoil cross sections. We have updated TALYS to estimate reaction rates of particular relevance to astrophysics. In contrast to other codes developed  for astrophysical applications, TALYS avoids many of the approximations mentioned above; it therefore provides a unique opportunity to test the robustness of these alternative codes. The principal advantages of the TALYS code and the new benefits that it provides in particular to he calculation of astrophysical reaction rates is the subject of the present paper.

In Sect.~\ref{sect_reacmod}, we summarize the reaction formalism and the major nuclear ingredients of relevance in reaction cross section calculations. In Sect.~\ref{sect_result}, we describe the advantages of the TALYS code. This includes a comparison with the code MOST, which has been widely used in astrophysical applications, in Sect.~\ref{sect_comp}. The pre-equilibrium process is shown in Sect.~\ref{sect_preeq} to have an important impact on the reaction rate. In Sect.~\ref{sect_n2n}, the (n,2n) reaction rate is calculated for the first time for all nuclei and its relevance is discussed. Finally, in Sect.~\ref{sect_concl}, conclusions are drawn.

\section{TALYS and nuclear reaction model}
\label{sect_reacmod}
Most nuclear astrophysical calculations adopt nuclear reaction rates evaluated
by the HF model.  It relies on the fundamental assumption (Bohr hypothesis) that the capture process occurs by means of the
intermediary production of a compound system that can reach a state of thermodynamic equilibrium. This compound system is then classically referred to as the compound nucleus (CN). The formation of  a CN occurs if the CN level density, at the excitation energy corresponding to the projectile incident energy, is sufficiently high.  The reaction
$I^{\mu} + a\rightarrow I' + a'$ represents the capture of a light particle $a$ onto a target $I$
in its state $\mu$, where $\mu=0$ corresponds to the target ground state, leaving the residual nucleus $I'$ and the particle or photon $a'$; TALYS estimates the corresponding cross section by the compound nucleus formula for the binary cross section in the full $jls$ scheme, i.e. 
\begin{eqnarray}
\sigma^{\mu}_{\alpha\alpha '} & = & D^{comp}\pi\lambdabar^{2} 
\sum_{J=mod(I^{\mu}+s,1)}^{l_{max}+I^{\mu}+s}\sum_{\Pi =-1}^{1}
\frac{2J+1}{(2I^{\mu}+1)(2s+1)} \cr
& & \sum_{j=|J-I^{\mu}|}^{J+I^{\mu}} \sum_{l=|j-s|}^{j+s} 
\sum_{j'=|J-I'|}^{J+I'} \sum_{l'=|j'-s'|}^{j'+s'}
 \delta_{\pi}(\alpha )\delta_{\pi}(\alpha ') \cr
 & &
\frac{\left< T_{\alpha lj}^{J}(E_{a})\right>
 \left< T_{\alpha 'l'j'}^{J}(E_{a'}) \right>}{\sum_{\alpha '',l'', j''}
\delta_{\pi}(\alpha '')
\left< T_{\alpha ''l''j''}^{J}(E_{a''}) \right>} W_{\alpha lj\alpha 'l'j'}^{J} ~ .
\label{eq_sig}
\end{eqnarray}
In the above equations,  $E_{a}$, $s$, $\pi_{0}$, $l$, and $j$ represent the projectile energy, spin, parity, orbital, and total angular momentum, respectively, and $l_{max}$ is the maximum projectile $l$-value. The same symbols but labelled by a prime correspond to the ejectile. The symbols $I^{\mu}$, $\Pi^{\mu}$ ($I'$, $\Pi '$) represent the spin and parity of the target (residual) nucleus, while $J$ and $\Pi$ correspond to the spin and parity of the compound system.  The initial system of projectile and 
target nucleus is designated by $\alpha=\{a,s,E_{a},E_{x}^{\mu},I^{\mu},\Pi^{\mu}\}$ 
where $E_{x}^{\mu}$ corresponds to the excitation energy of the 
target nucleus. $\alpha '=\{a',s',E_{a'},E_{x}',I',\Pi '\}$ is similar for the final system of ejectile and 
residual nucleus. $\delta_{\pi}(\alpha )=1$, if $(-1)^{l}\pi_{0}\Pi^{\mu}=\Pi$ and 0 
otherwise (the same holds for the final system $\alpha '$). In Eq.~\ref{eq_sig}, $\lambdabar$ is the relative motion wave length, $T$ the transmission coefficient, $W$ the  width fluctuation correction factor and $D^{comp}$ the depletion factor  given by 
\begin{equation}
D^{comp}=[\sigma_{reac}-\sigma^{disc,direct}-\sigma^{\rm PE}]/\sigma_{reac}
\label{eq_d}
\end{equation}
where $\sigma_{reac}$ is 
the total reaction cross section, $\sigma^{disc,direct}$ is the total discrete direct cross section, and $\sigma^{\rm PE}$ is the pre-equilibrium cross section.
It is assumed by the TALYS code that direct and compound contributions can be added incoherently. The 
formula for $D^{comp}$ is only applied to weakly coupled channels that deplete 
the flux, such as contributions from the direct or pre-equilibrium processes. For deformed nuclides, the effect of direct transitions to discrete levels is included directly in the coupled-channels scheme and the $\sigma^{disc,direct}$ term is omitted from Eq.~\ref{eq_d}.

The HF formalism is valid only if the formation and decay of the CN are totally independent. This
so-called Bohr hypothesis may not be fully satisfied, particularly in cases where a few
strongly and many weakly absorbing channels are mixed. As an example, the HF equation is
known to be invalid when applied to the elastic channel, since in that case the
transmission coefficients for the entrance and exit channels are identical, and hence
correlated. These correlations enhance the elastic channel and accordingly decrease the other open channels. To account for these deviations, a width
fluctuation correction $W$ is introduced into the HF formalism (see Eq.~\ref{eq_sig}).  The different approximate expressions included in TALYS are described and discussed in Hilaire et al. (2003). When many competing channels are open, above a few MeV of projectile energy,  the width fluctuation correction can be neglected.
 
Each transmission coefficient $T$ is estimated for all levels with
experimentally known energy, spin, and parity. In that case, we simply have $\left<T_{\alpha 'l'j'}^{J}(E_{a'})\right> = T_{\alpha 'l'j'}^{J}(E_{a'})$. However, if the excitation energy $E_{x}'$, which is implicit in the definition of channel  $\alpha '$, corresponds to a state in the continuum, we have an 
effective transmission coefficient for an excitation energy bin of 
width $\Delta E_{x}$ determined by the integral 
\begin{equation}
\left<T_{\alpha 'l'j'}^{J}(E_{a'})\right> =
\int_{E_{x}'-\frac{1}{2}\Delta E_{x}}^{E_{x}'+\frac{1}{2}\Delta E_{x}} dE_{y} 
\rho (E_{y},J,\Pi ) T_{\alpha 'l'j'}^{J}(E_{a'})
\label{eq_trans}
\end{equation}
over the level density $\rho (E_{y},J,\Pi )$, at an energy $E_y$, spin $J$, and parity $\Pi$ in the CN or residual nucleus.

 For increasing energy or nuclei for which the CN does not have time to reach thermodynamic equilibrium, direct or pre-equilibrium processes may become significant.  In TALYS, the direct component is derived from  the Distorted Wave Born Approximation  for spherical nuclei and the coupled-channels equations for deformed nuclei. The pre-equilibrium emission can occur after the first stage of the reaction but long before statistical equilibrium of the CN is reached.  Although pre-equilibrium processes can cover a sizable part of the reaction cross section for intermediate energies (typically above a few MeV in stable nuclei), they have always been neglected to estimate the reaction rates for astrophysical applications.  Both classical and quantum-mechanical pre-equilibrium models exist and are included in TALYS. One of the most widely used pre-equilibrium models is the (one- or two-component) exciton model (see Koning \&Duijvestijn 2004 for an extensive  review),  in which the nuclear state is characterized at any moment during the reaction 
by the total energy $E^{tot}$ and the total 
number of particles above and holes below the Fermi surface. Particles ($p$) 
and holes ($h$) are referred to as excitons. Furthermore, it 
is assumed that all possible ways of sharing the excitation energy between 
different particle-hole configurations at the same exciton number $n=p+h$ 
have an equal a-priori probability.
To monitor the evolution of the scattering process, one merely traces
the temporal development of the exciton number, which changes in time as a
result of intranuclear two-body collisions. The basic starting point of the
exciton model is a time-dependent master equation, which describes the 
probability of transitions to more and less complex
particle-hole states as well as transitions to the continuum (emission).
Upon integration over time, the energy-averaged emission spectrum is derived.  
These assumptions ensure that the exciton model is amenable to practical calculations.
The disadvantage, however, is the introduction of a free parameter,
namely the average matrix element of the residual two-body interaction,
occurring in the transition rates between two exciton states. 

A thermodynamic equilibrium holds locally to a good approximation inside stellar interiors. Consequently, the energies of both the targets and projectiles, as well
as their relative energies $E$, obey a Maxwell-Boltzmann distribution corresponding to the
temperature $T$ at that location (or a blackbody Planck spectrum for photons). 
In such conditions, the astrophysical rate is obtained by integrating the
cross section given by Eq.~(\ref{eq_sig}) over a Maxwell-Boltzmann distribution of energies
 at the given temperature $T$. In addition, in hot astrophysical plasmas, the target nucleus
exists in its ground as well as excited states. In a thermodynamic equilibrium situation, 
the relative populations of the various levels of nucleus $I^\mu$ with excitation energies
$E_x^{\mu}$ obey a Maxwell-Boltzmann distribution.
The effective stellar rate  per pair of particles in the entrance
channel at a temperature $T$, taking account of the contributions of various target
excited states, is finally expressed in a classical notation as
%--------------------------------------------------------------------
\begin{eqnarray}
 & &N_A\langle\sigma v\rangle^*_{\alpha \alpha '}(T) = \Bigl( \frac{8}{\pi m} \Bigr) ^{1/2}
                  \frac{N_{A}}{(kT)^{3/2}~G_I(T)}\cr
              & & ~   \int_{0}^{\infty} \sum_{\mu} \frac{(2I^{\mu}+1)}{(2I^{0}+1)} 
                    \sigma^{\mu}_{\alpha\alpha '}(E)E\exp \Bigl(-\frac{E+E_x^{\mu}}{kT}\Bigr) dE,
\label{eq_rate}
\end{eqnarray}
%--------------------------------------------------------------------
%
where $k$ is the Boltzmann constant,
$m$ the reduced mass of the $I^0 + a$ system, $N_{A}$ the Avogadro number, and
\begin{equation}
G_I(T) =\sum_{\mu} {(2I^{\mu}+1)}/{(2I^{0}+1)}\exp(-E_x^{\mu}/kT)
\end{equation}
 the  $T$-dependent normalized partition function. 
 
 Reverse reactions can also be estimated by
making use of the reciprocity theorem (Holmes et al. 1976). In particular, the stellar
photodissociation rates for astrophysical applications have until now been derived entirely from radiative capture rates by the expression
%
%--------------------------------------------------------------------
\begin{eqnarray}
\lambda_{(\gamma,a)}^*(T) = & & 
\frac{(2I^{0}+1)(2j_a+1)}{(2I'^{0}+1)}~\frac{G_I(T)}{G_{I'}(T)}
 \Bigl( \frac{m kT}{2 \pi \hbar^2} \Bigr)^{3/2} \times \cr 
 & & \langle\sigma v\rangle^*_{(a,\gamma)}
~ {\rm e}^{-Q_{a\gamma} /kT},
\label{eq_inverserate}
\end{eqnarray}
%--------------------------------------------------------------------
%
where $Q_{a\gamma}$ is the Q-value of the $I^0(a,\gamma)I'^0$ capture. We note that,
in stellar conditions, the reaction rates for targets in thermal equilibrium are usually believed to 
obey reciprocity since the forward and reverse channels are expected to be symmetrical, in contrast to the
situation that would be experienced by targets in their ground state only (Holmes et al. 1976). 
The total stellar photodissociation rate can also be determined directly from 
%
%--------------------------------------------------------------------
\begin{equation}
\lambda^*_{(\gamma,a)}(T) = \frac
{\sum_\mu (2I^\mu+1)~\lambda^{\mu}_{(\gamma,a)}(T)~ \exp(-E_x^{\mu}/kT)}
{\sum_\mu (2I^\mu+1)~ \exp(-E_x^{\mu}/kT)},
\label{eq_photorate}
\end{equation}
%--------------------------------------------------------------------
%
where the photodissociation rate $\lambda^{\mu}_{(\gamma,a)}$ of state $\mu$ with excitation
energy $E_x^{\mu}$ is given by 
\begin{equation}
\lambda^{\mu}_{(\gamma,a)}(T) = \int^\infty_0 c~n_\gamma(E,T)~\sigma^{\mu}_{(\gamma,a)}(E)\,dE\,,
\label{eq_gn}
\end{equation}
%---------------------------------------------------
%
where $c$ is the speed of light, $\sigma^{\mu}_{(\gamma,a)}(E)$ is the photodisintegration
cross section at energy $E$ estimated within the above-described reaction formalism, and $n_\gamma$ is the stellar $\gamma$-ray distribution described by the blackbody Planck spectrum at the given temperature $T$.

The uncertainties involved in any cross section calculation are of two origins:
\begin{itemize}
\item 
the first one is related to the reaction mechanism, i.e. the model of formation and de-excitation of the CN itself. Reaction mechanisms have compound, pre-equilibrium, and direct components. The compound formation is described by the HF theory. The pre-equilibrium and direct contributions are often neglected, in particular in astrophysical codes. The impact of this non-equilibrium component is studied in Sect.~\ref{sect_preeq}. More specifically, the predictions derived by the TALYS  and MOST (Aikawa et al. 2005, Arnould \& Goriely 2006) codes are compared in Sect.~\ref{sect_comp} by applying the same nuclear input models. The added-values of TALYS are further discussed in Sect.~\ref{sect_result}. 

\item 
another type of uncertainty comes from the evaluation of the nuclear quantities required for
calculating the transmission coefficients in  Eqs.~(\ref{eq_sig}) -
(\ref{eq_rate}), i.e. the ground and excited state properties (masses,
deformations, matter densities, excited state energies, spins, and parities, \dots), nuclear level densities, $\gamma$-ray strength, optical model potential, and fission properties.   When not
available experimentally, this information has to be derived from nuclear models. 
\end{itemize}

Ideally, when dealing with nuclear astrophysics applications, the various nuclear ingredients should be determined from {\it global}, {\it universal}, and {\it microscopic} models. 
The large number of nuclides involved in the modeling of some nucleosynthesis mechanisms implies that global models should be used. On the other hand, a universal description of all nuclear properties within a unique framework for all nuclei involved in a nuclear network ensures the essential
coherence of predictions for all unknown data. Finally, a microscopic description provided
by a physically sound theory based on first principles ensures  extrapolations away from
experimentally known energy or mass regions that are likely to be more reliable than
predictions derived from more or less parameterized approaches of various types and
levels of sophistication.
This is true even as new generations of such models are starting to be developed to compete with more phenomenological  highly-parameterized models for the reproduction of experimental data  (Goriely et al. 2002, 2004, 2007; Hilaire \& Goriely 2006). 
Only a few reaction model codes (for astrophysical applications) adopt the largest possible extent of global and coherent microscopic (or at least semi-microscopic) models. This is the case of MOST (Aikawa et al. 2005, Arnould \& Goriely 2006) and now TALYS, but this is not the case of neither the original work of Holmes et al. (1976), nor the code SMOKER (Thielemann et al. 1987) nor  NON-SMOKER (Rauscher \& Thielemann 2001).  We note that  MOST, SMOKER and NON-SMOKER codes are similar (if not identical) in their description of the reaction mechanisms and corresponding approximations, but adopt different  nuclear input models.

It should be mentioned that TALYS includes the possibility to calculate reaction rates with both different input parameters and a number of alternative local or global models for the various nuclear ingredients. This feature allows the possible study of  not only parameter uncertainties (and therefore covariance matrices), but also model uncertainties. The readers are referred to Aikawa et al. (2005), Arnould\& Goriely (2006) and Arnould et al. (2007)  for a discussion on how nuclear models affect the reaction rates of relevance to nucleosynthesis.

  Finally, we note that TALYS, in a similar way to  all the astrophysically orientated codes,  does not describe the direct {\it capture} mechanism (Christy \& Duck 1961; Satchler 1983), which is known to play an important role in light, closed shell, or heavy exotic neutron-rich systems for which no resonant states are available. The contribution of  the direct capture of heavy exotic neutron-rich nuclei has been taken into account only by a few restricted studies (Mathews et al. 1983; Goriely 1997, 1998; Rauscher et al. 1998). This direct radiative capture is therefore not included in TALYS.  The so-called direct process in TALYS corresponds to the elastic or inelastic excitation of the discrete levels at the highest outgoing energies, and is described using either the distorted wave Born approximation for spherical nuclei or the coupled-channel equations for deformed nuclei. The low-energy direct radiative capture to a final discrete state, as described for example by the potential model (Goriely 1997) and  involving the overlap of the initial scattering wave function in the entrance channel, the bound-state wave function and the electromagnetic operator, is not taken into account by the TALYS code, and is  not discussed further here.
We do mention however that a simple but powerful simulation of the direct-semidirect capture process
is included in TALYS by  the exciton model (Akkermans \& Gruppelaar, 1985).

\section{The advantage of TALYS predictions}
\label{sect_result}
Concerning the calculation of Maxwellian-averaged reaction rates of astrophysical interest, the TALYS code has some clear advantages over previous codes developed for astrophysical applications (like MOST), such as 
\begin{itemize}
\item the inclusion of the pre-equilibrium reaction mechanism  (totally neglected in other astrophysical nuclear codes), 
\item  the detailed description of the decay scheme, including the description of $\gamma$-delayed particle emission and the possible particle emission from all residual nuclei. In astrophysical codes, the particle emission following a primary $\gamma$ cascade is not explicitly taken into account, but corrected roughly by artificially reducing the photon transmission coefficient in the numerator of Eq.~\ref{eq_sig}, as described by  Holmes et al. (1976);
\item the  inclusion of multi-particle emission (totally neglected in other astrophysical nuclear codes);
\item the inclusion of detailed width fluctuation corrections (Moldauer 1980, Verbaarschot 1985) in contrast to the approximation of  Hofmann et al. (1980) applied in other astrophysical codes;
\item the inclusion of parity-dependent level densities  (the parity equipartition is usually assumed in astrophysical codes, one exception being the work of  Mocelj et al. 2007, which however considers a spin-independent parity distribution); 
\item the inclusion of a coupled channel description for deformed nuclei,  while the other astrophysical codes consider spherically equivalent optical potentials for deformed targets, and 
\item   the inclusion of the fission channel for the compound as well as the residual nuclei. Fission is often neglected in astrophysical codes, and if included, insufficiently tested on experimental data and not consistently taken into account in the full decay scheme. 
\end{itemize}
%---------------------------------------------------
\begin{figure}
\centering
\includegraphics[scale=0.31]{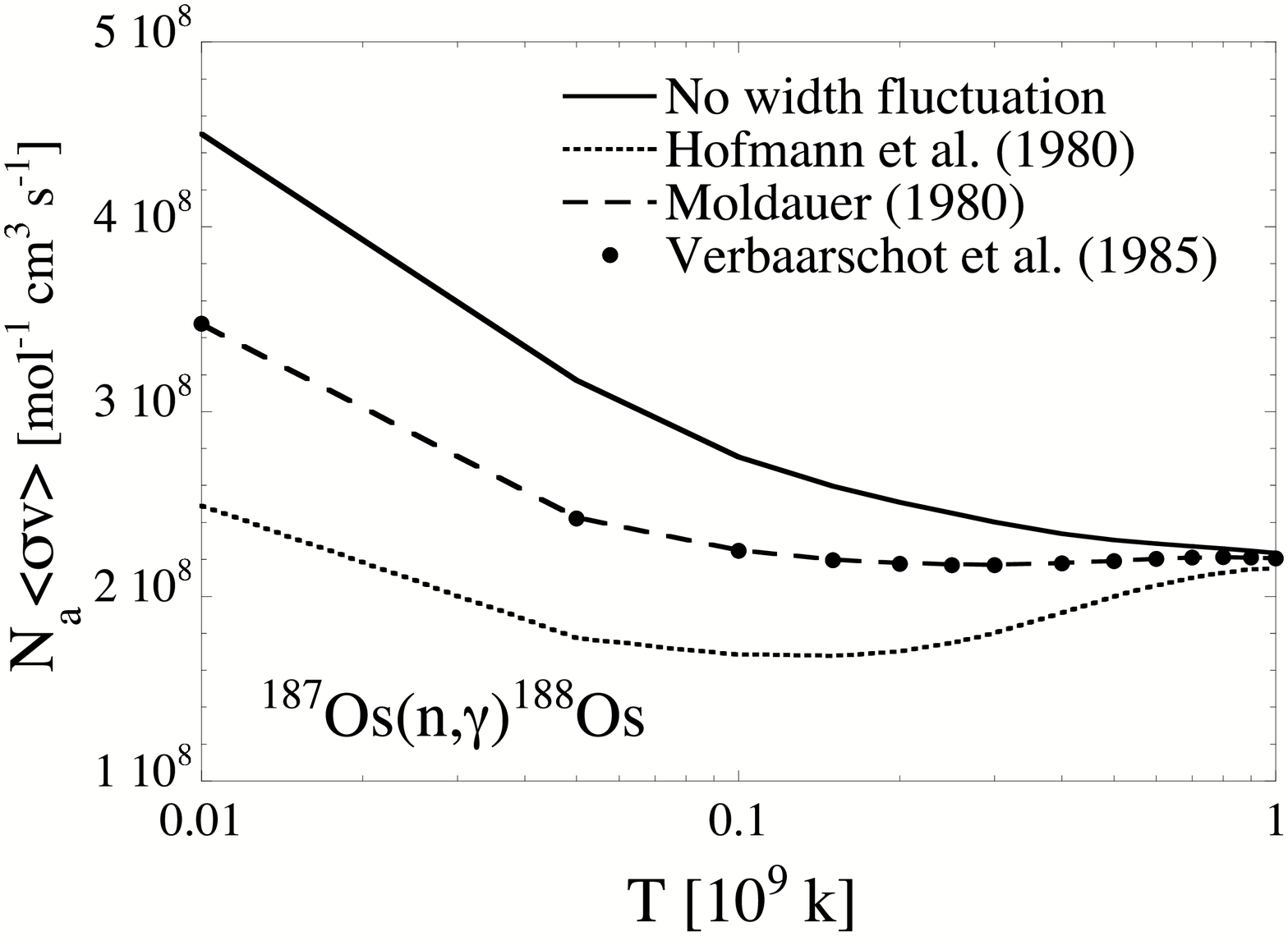}
\vskip 1cm
\caption{Stellar $^{187}$Os(n,$\gamma$)$^{188}$Os rate as a function of the temperature for 3 different prescriptions of the width fluctuation correction and without any. }
\label{fig01}
\end{figure}
%---------------------------------------------------

For the calculation of thermonuclear reaction rates at astrophysically relevant temperatures, most approximations assumed by previous codes were justified, since for neutron captures only incident energies of a few keV to 1 MeV contribute, and for incident charged particles only energies below or close to the Coulomb barrier come into play. Most of the above mentioned {\it ``added values"} therefore do not affect the predicted rates by more than a few tens of a percent and remain insignificant with respect to the uncertainties in the nuclear ingredients, such as nuclear properties, level densities, or $\gamma$-ray strengths.  This concerns in particular the influence of effects such as multi-particle emission, the width fluctuation correction, the parity-dependent level density or the coupled channel description for deformed nuclei. Their impact on astrophysics applications may consequently be relatively limited.

However, in some specific astrophysical applications, highly accurate predictions may be required. This is the case for the Re-Os nucleocosmochronology, which is known to be sensitive to the estimate of the $^{187}$Os(n,$\gamma$)$^{188}$Os reaction rate and most particularly to the  experimentally unknown contribution of the thermally populated 9.75keV first excited state of $^{187}$Os. A significant amount of  experimental data were gathered to constrain the determination of this reaction rate directly or indirectly (see e.g. Segawa et al. 2007 and references therein). In this specific example, to reconcile all of the direct and indirect constraints obtained experimentally, the state-of-the-art reaction model should be used, and in  particular a proper description of the width fluctuations is required  (Hilaire et al. 2003). The TALYS code enables the neutron capture rates to be estimated with an improved description of the width fluctuation correction based in particular on the exact, but very computer-time-consuming, Gaussian Overlap Ensemble (GOE) approach of Verbaarschot et al. (1985) or its excellent approximation due to Moldauer (1980).  Figure~\ref{fig01} shows that the correction of Hofmann et al. (1975, 1980), traditionally adopted in all previous works related to the Re-Os chronology, differs  by some 40\% from the GOE predictions  at the relevant low temperatures of $1-3 \times 10^8$~K.  Such a difference significantly affects the determination of the stellar $^{187}$Os(n,$\gamma$)$^{188}$Os reaction rate, and, consequently, the determination of the age of the Galaxy.

In addition to this improved modeling of the width fluctuation correction, TALYS also enables us to test for the first time some of the approximations made in astrophysical codes, as shown below.  It is impossible to describe all nuclear astrophysical applications and to fully assess the impact the newly-derived TALYS rates can have.  For this reason,  we restrict ourselves to a comparison of the TALYS predictions with those derived by more traditional Hauser-Feshbach codes such as MOST; this demonstrates the extent to which some of the above-listed added values can affect {\it globally} the reaction rate predictions and potentially some nucleosynthesis processes. 

We note that we do not study the fission channel. TALYS was extensively used in data evaluation for nuclear power applications and the fission channel was tested on experimental data in far greater detail than for other astrophysical codes (see for example, Duijvestijn \& Koning 2005). In particular, it should be mentioned that TALYS considers deformed optical model potentials and can account
for the calculation of the penetration through a numerical non-parabolic barrier obtained from
HFB calculations using the corresponding HFB plus combinatorial NLD at the saddle points (Sin et al. 2007; Goriely et al. 2008).
Finally, TALYS takes into account the possible effect of discrete transition and class II states that may affect the reaction cross section at low energies. However, with TALYS like any other code, it remains difficult to predict (with a satisfactory accuracy) fission cross sections on the basis of global input models. For this reason, a detailed study dedicated to fission will be presented in a future specific publication (Goriely et al. 2008).

\subsection{Comparison between MOST and TALYS predictions}
\label{sect_comp}

The present section aims to compare TALYS predictions with those derived by the MOST code which has been widely used in astrophysical applications (Aikawa et al. 2005; Arnould \& Goriely 2006).   The same nuclear input models are adopted for the comparison, namely  the Hartree-Fock-Bogolyubov (HFB)  nuclear masses  (Goriely et al. 2007), the HFB plus combinatorial nuclear level density (Hilaire \& Goriely 2006), the HFBCS plus QRPA $\gamma$-ray strength (Goriely \& Khan 2002), the global nucleon optical model potential of Koning \& Delaroche (2003), and the $\alpha$-potential of McFadden \& Satchler (1966).  Whenever available, experimental data constraining such models are included, although both reaction codes do not make use of exactly the same renormalization procedures.  It should be noted that an accurate comparison of both reaction codes remains impossible due to the different recipes used either in the renormalization procedure or the way in which experimental data are used explicitly in the code or implicitly to constrain theoretical inputs. In particular, the number of discrete states, as well as their spin and parity assignment could differ. TALYS code also renormalizes the $\gamma$ strength function on radiative strength at the neutron binding energy when available. This step is not included by MOST. 

\begin{figure}
\centering
\includegraphics[scale=0.32]{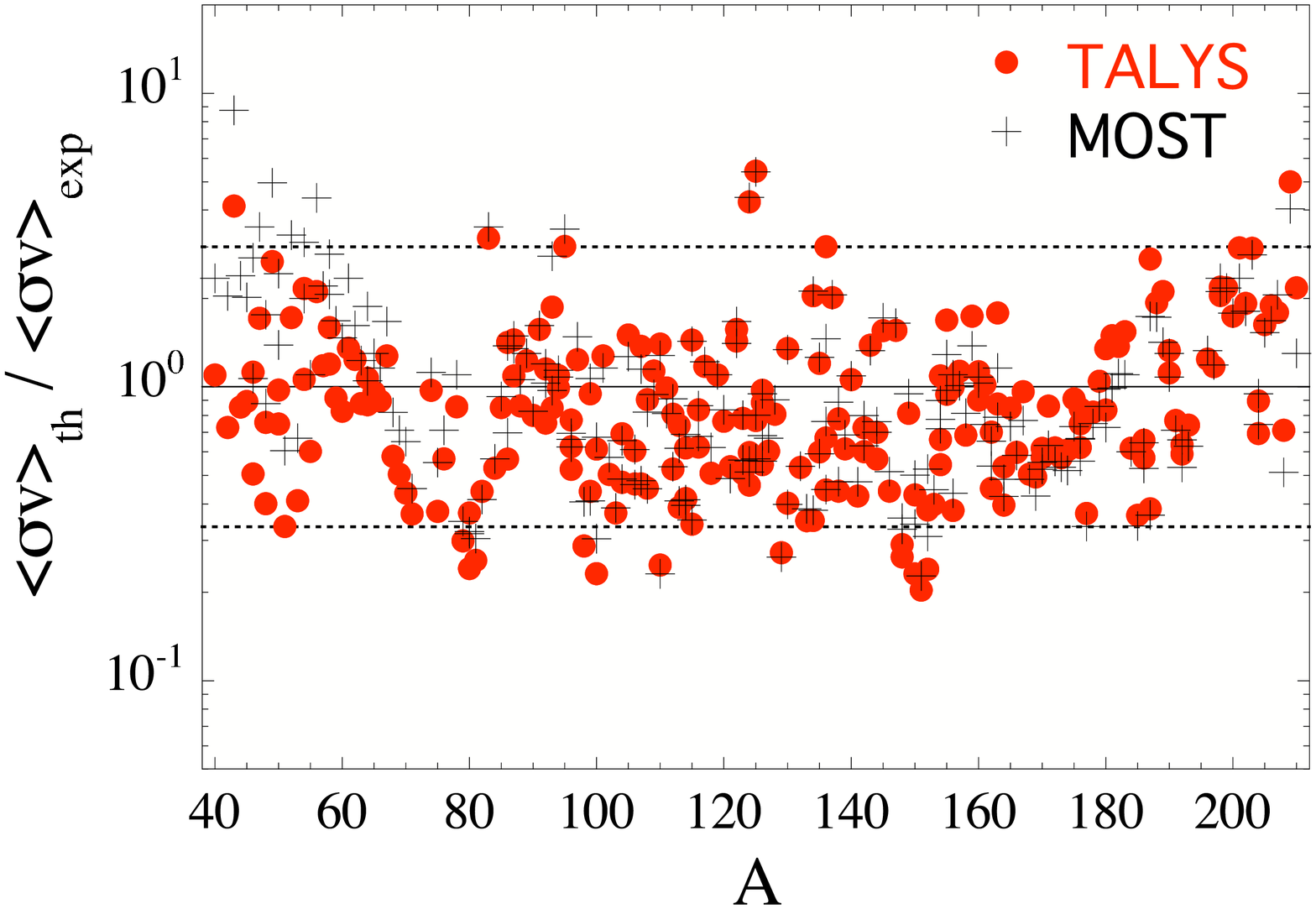}
\vskip 1cm
\caption{Comparison of MOST (crosses) or TALYS (circles) radiative neutron capture rates with experimental data above Ca (Bao et al. 2000) at the temperature $T=3~10^8$~K. }
\label{fig02}
\end{figure}

With such global inputs, both codes reproduce the experimental neutron, proton or photo-rates with the same accuracy, i.e. to within roughly a factor of 2--3, as shown in Fig.~\ref{fig02} for the radiative neutron capture rates of particular interest in s-process nucleosynthesis studies.  Differences can however be noted. These arise partly because of the different databases of experimental structure properties that were adopted and more particularly the way in which the deformation effects were treated. In MOST (but not in TALYS), the Fe-group nuclei are described as rotationally deformed nuclei rather than vibrational nuclei. This explains the larger neutron capture rate of low-mass nuclei. TALYS also estimates the width fluctuation correction on the basis of the Moldauer (1980) prescription, which tends to increase (see Fig.~\ref{fig01}) the reaction cross section relative to the Hofmann et al. (1980) correction adopted in MOST. This effect is stronger in spherical nuclei and explains the tendency for TALYS to predict larger neutron rates than MOST for heavy-mass nuclei.

%---------------------------------------------------
%---------------------------------------------------

\begin{figure}
\centering
\includegraphics[scale=0.32]{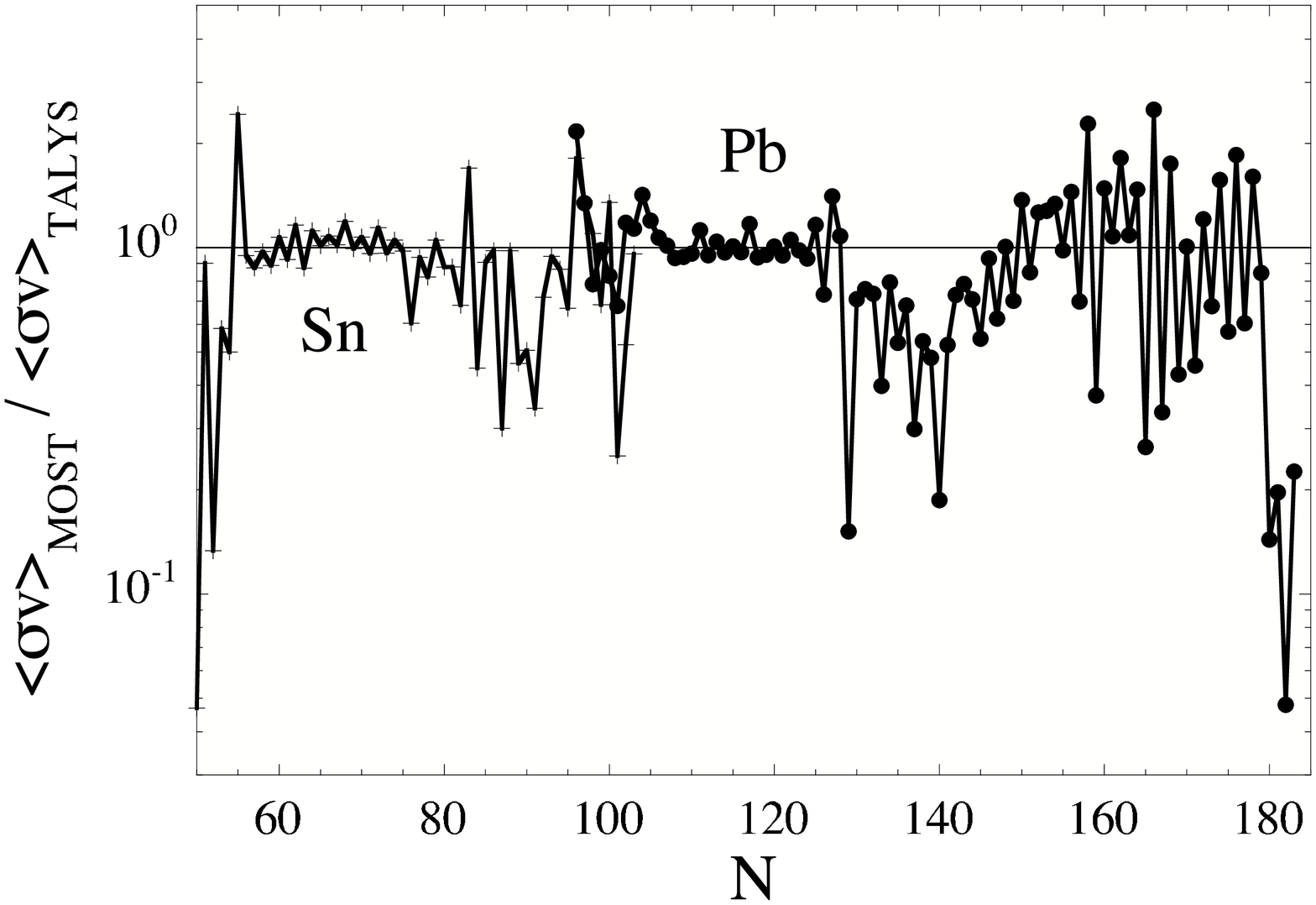}
\vskip 1cm
\caption{Ratio of MOST to TALYS radiative neutron capture rates for all Sn (crosses) and Pb (circles) isotopes at $T=10^9$~K as a function of the neutron number $N$. }
\label{fig03}
\end{figure}

\begin{figure}
\centering
\includegraphics[scale=0.32]{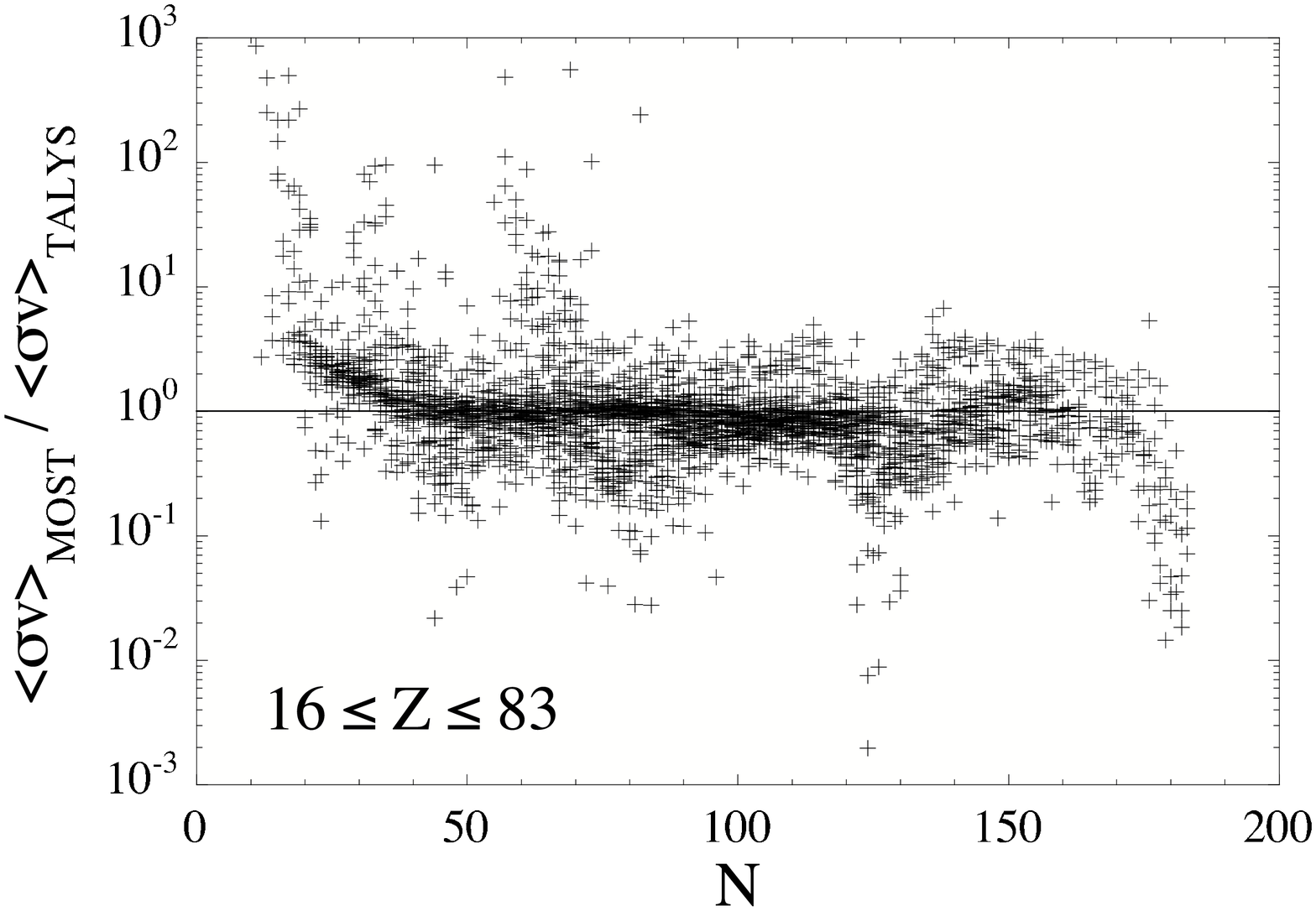}
\vskip 1cm
\caption{Ratio of MOST to TALYS radiative neutron capture rates for all nuclei with $16 \le Z \le 83$ at $T=10^9$~K as a function of the neutron number $N$. }
\label{fig04}
\end{figure}

Although some discrepancies can be observed, both codes predict globally very similar radiative neutron capture rates for the stable nuclei. This may not be the case for exotic neutron-rich nuclei, such as those involved in the r-process nucleosynthesis  (for a review on the r-process, see Arnould et al. 2007). We show in Figs.~\ref{fig03}--\ref{fig04} the discrepancies that could exist between MOST and TALYS estimates of the radiative neutron capture rates for the Sn and Pb isotopes (Fig.~\ref{fig03}) and more generally for all nuclei with $16 \le Z\le 83$ between the neutron and proton drip lines (Fig.~\ref{fig04}). This comparison is made for a temperature of $T_9=1$ (where $T_9$ is the temperature expressed in $10^9$~K), which is typical of conditions expected during the neutron captures by exotic neutron-rich nuclei in the r-process nucleosynthesis. As long as we deal with stable nuclei, both codes predict almost identical rates. However, deviations by a factor as large as 10 for Sn and Pb are seen, as soon as nuclei away from the valley of $\beta$-stability are considered. 
 
These deviations can be understood in terms of 
\begin {itemize}
\item the sensitivity of the radiative neutron capture, especially for nuclei with low neutron separation energy ($S_n$), to the nuclear structure properties and nuclear ingredients entering the reaction model. In particular, such a comparison emphasizes the difficulty in predicting neutron-capture cross sections when no or little experimental data are available to constrain the nuclear inputs or to enter directly into the calculation. For radiative neutron capture, this is particularly the case for spherical targets (the HFB mass model used here, namely HFB-14, predicts the $142 \la A \la 163$ Sn and $223 \la A \la 248$ Pb n-rich isotopes to be deformed). For example, the  $^{211}$Pb neutron capture cross section and rate differ by about a factor of 7 (Fig.~\ref{fig03}). This deviation is essentially due to the number of experimental discrete levels of the compound nucleus included in each calculation. Different prescriptions are used to estimate the number of levels up to which the level scheme is believed to be complete and, in this specific case, this has a direct impact on the low-energy cross section.

\item the reaction mechanism and its particular implementation.  In particular,  the contribution of the pre-equilibrium process, totally neglected in other astrophysical codes, in fact significantly modifies the cross section of the keV-MeV region of astrophysical interest (as described below). This pre-equilibrium effect is also responsible for the odd-even staggering observed for exotic neutron-rich nuclei in Fig.~\ref{fig03} (as discussed in Sect.~\ref{sect_preeq}). 
\end{itemize}

%---------------------------------------------------
%---------------------------------------------------
%---------------------------------------------------
%---------------------------------------------------

\subsection{The pre-equilibrium contribution}
\label{sect_preeq}
\begin{figure}
\centering
\includegraphics[scale=0.42]{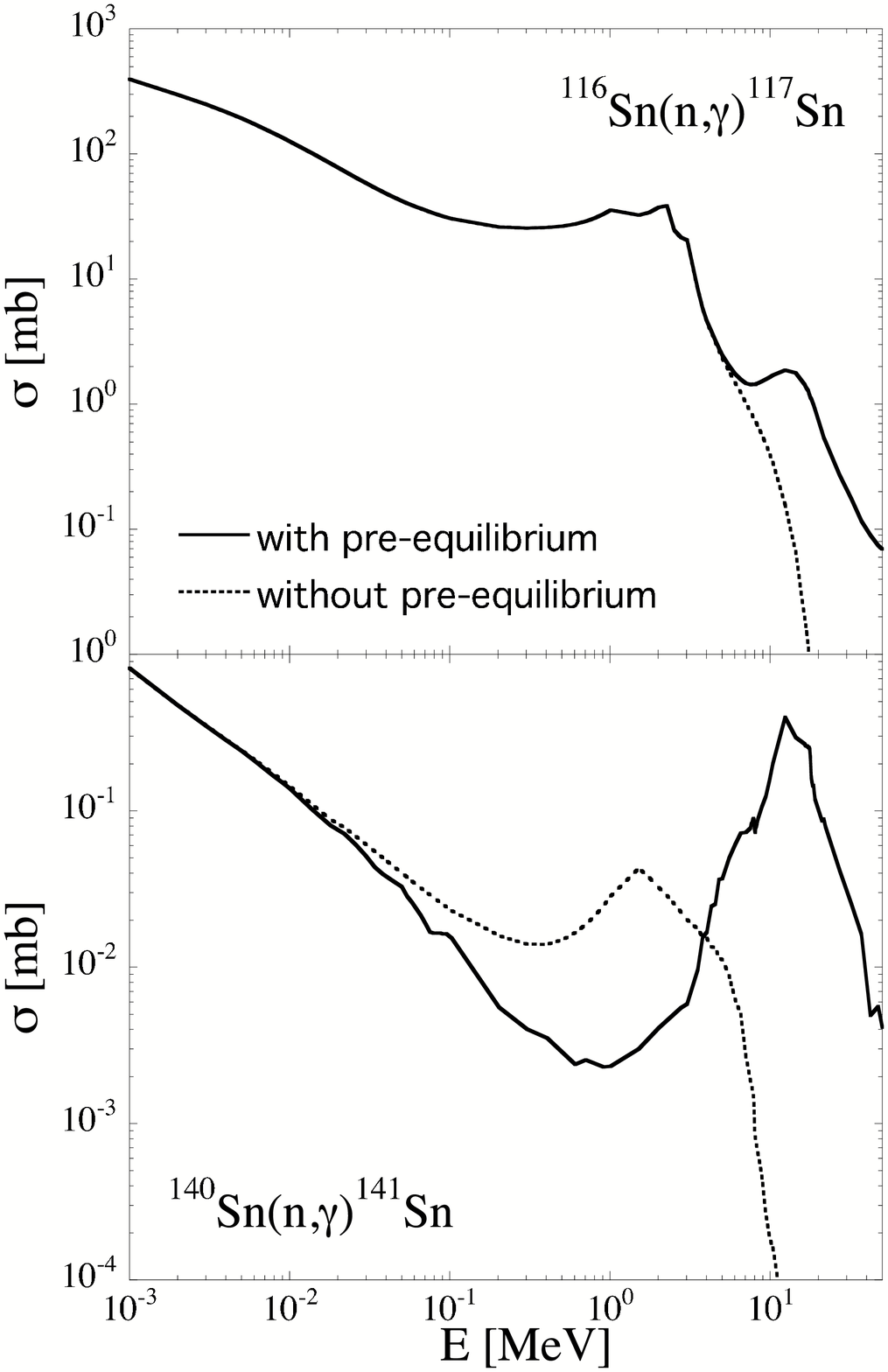}
\caption{Radiative neutron capture cross section for $^{116}$Sn (top) and $^{140}$Sn (bottom) with and without including the pre-equilibrium processes as a function of energy}
\label{fig05}
\end{figure}

\begin{figure}
\centering
\includegraphics[scale=0.42]{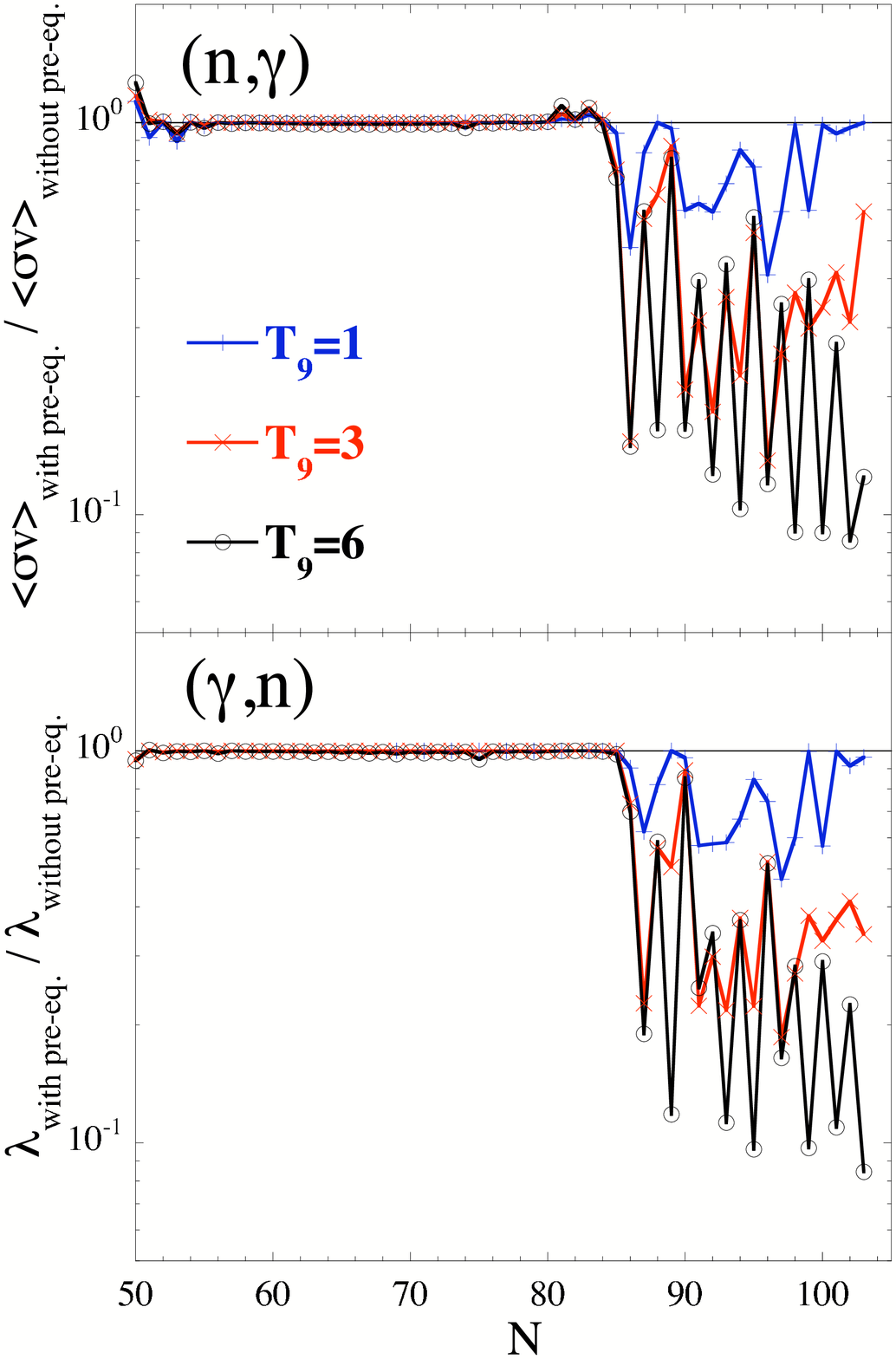}
\caption{Ratio of the radiative neutron capture (top) and photoneutron (bottom) rates obtained with and without including the pre-equilibrium processes for all Sn isotopes at 3 different temperatures $T_9$.}
\label{fig06}
\end{figure}

In the empirical picture of a nuclear reaction, it is imagined that the incident particle step-by-step creates more complex states of excited particles and holes in the compound system of target plus projectile and thereby gradually loses its memory of the initial energy and direction.
Pre-equilibrium emission occurs after the first stage of the reaction but long before statistical equilibrium of the CN is achieved, leading to the well-known high-energy tails in the emission spectra and the smooth forward peaked angular distributions.  The pre-equilibrium process is also responsible for the observed increase in the radiative neutron-capture cross section for stable nuclei at incident energies typically around 10~MeV (see e.g. upper panel of Fig.~\ref{fig05}), which hardly affects the Maxwellian-averaged reaction rates of astrophysical relevance. However, if we consider exotic neutron-rich nuclei with low neutron separation energies, the low-level density makes it difficult for the nucleus to achieve complete equilibrium. In this case, the pre-equilibrium process starts to affect the neutron channel already at a few hundred keV, and consequently may change the astrophysical rate. An example is shown in Fig.~\ref{fig05}, where it can be seen that already at 100~keV, the pre-equilibrium process affects the capture cross section of  $^{140}$Sn, while for stable nuclei, such as $^{116}$Sn, it starts to contribute only above some 7~MeV.  Interestingly, for neutron-rich nuclei the calculated capture cross section for the keV-MeV range is smaller when the pre-equilibrium mechanism is included rather than excluded. This is because  already at these low energies the neutron inelastic channel basically exhausts the
total non-elastic cross section. 
The level density and optical models, steering the reaction processes at these energies, determine that the relatively small contribution of capture, in contrast to inelastic scattering, is lower for pre-equilibrium processes than for compound processes. We note that the pre-equilibrium model adopted here corresponds to the two-component exciton model (Koning \& Duijvestijn 2004), which remains relatively phenomenological as discussed in Sect.~\ref{sect_reacmod}. Its reliability and quantitative contribution, when applied to neutron-rich nuclei, can therefore be questioned. However,  such an effect is found to be qualitatively non-negligible for exotic neutron-rich nuclei of small neutron separation energies, regardless of the nuclear input adopted.  The decrease in the reaction cross section for the 100~keV region affects directly the Maxwellian-averaged reaction rates of relevance to astrophysics as seen in Fig.~\ref{fig06} (upper panel). At increasing temperatures, the Gamow energy increases and overlaps the energy region in which the pre-equilibrium process contributes to the reaction mechanisms. The pre-equilibrium correction is found to be less important for odd-N target nuclei than for even-N. This is essentially due to the increase in the HF cross section for the 0.5--2~MeV region found for even-N target nuclei, which results from the oddness of the compound system that has a larger nuclear level density and the evenness of the target reducing the inelastic channel competition. 

The pre-equilibrium contribution is found to decrease not only the radiative capture, but also the (n,2n) reaction rates of neutron-rich nuclei at temperatures around $T\simeq 2-4 \times 10^9$~K by about 50\%. It affects not only the nucleon or $\alpha$-particle capture rates but also the photoneutron rates of neutron-rich nuclei, as illustrated in Fig.~\ref{fig06} (lower panel). Since the pre-equilibrium component contributes increasingly more as energies increase, the higher the temperature the larger the impact on the astrophysics rates, as seen in Fig.~\ref{fig06}. The impact is also observed to be stronger for odd-N compound nuclei (i.e. for the neutron capture by an even-N target or the photoneutron emission of an odd-N target).

\subsection{The multi-particle emission}
\label{sect_n2n}

All reaction codes dedicated to astrophysical applications have made the simplified assumption that the contribution of the emission of more than one particle  in neutron or charged-particle induced reactions is negligible. In most astrophysical scenarios, the temperatures are indeed sufficiently low for such channels not to compete. However, it remained impossible to test such an approximation, because until now, no reaction code for astrophysical applications has included the multi-particle emission in their framework. By applying the TALYS code, it is  possible to have a more robust understanding of the competing channels that until now have been neglected. An important one is the (n,2n) channel that may dominate, at relatively high temperatures, the (n,$\gamma$) channel, especially for exotic neutron-rich nuclei. For energies above the two-neutron separation energies, the (n,2n) channel is open and begins to dominate very quickly, so that at a given temperature the ratio $r_n=\langle \sigma v \rangle_{(n,2n)}/\langle \sigma v \rangle_{(n,\gamma)}$ becomes larger than 1. This temperature is critical in applications such as the r-process nucleosynthesis, since above such a temperature, the neutron capture will not lead to the production of heavier neutron-rich nuclei, but in contrast will start to reverse the nuclear flow and prevent it from reaching the neutron-rich region.  

\begin{figure}
\centering
\includegraphics[height=9cm,width=9cm]{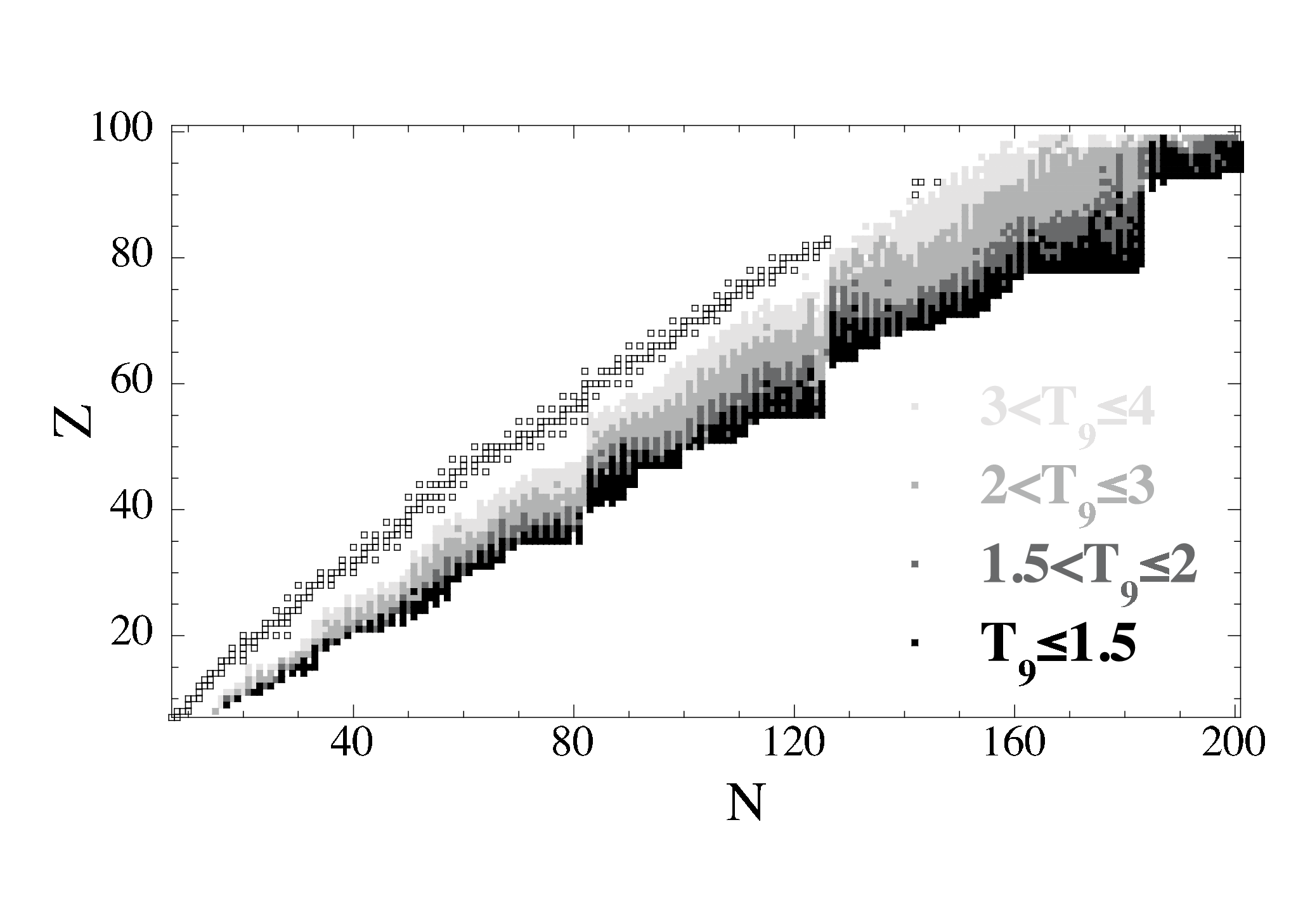}
\vskip 0.5 cm
\caption{Representation in the (N,Z) plane, as given by the legend, of the value of the critical temperature at which the Maxwellian-averaged (n,2n) rate becomes faster than the (n,$\gamma$) rate, i.e $r_n=1$.}
\label{fig08}
\end{figure}

In Fig.~\ref{fig08}, we show  the evolution of the critical temperature in the (N,Z)-plane. It can be seen that at temperatures above $3\times 10^9$~K, most of the neutron-rich region cannot be reached by neutron captures {\it regardless} of the neutron density in the astrophysical environment. At $T_9=3$ for example, within the Cd or Sn isotopic chain, the r-process flow could not reach isotopes heavier than $^{131}$Cd or $^{133}$Sn, respectively. At such high temperatures, the (n,2n) channel represents a nuclear barrier that the r-process flow would not be able to cross,  independently of the amplitude of the neutron density. The nuclear flow into the neutron-rich region is usually believed to be stopped by photodissociation. However, this $(\gamma,n)$ barrier at a given temperature can always be counterbalanced by considering higher neutron densities, while the $(n,2n)$ competition represents a barrier to the $(n,\gamma)$ flow independent of the neutron density. This nuclear barrier could significantly modify the prediction of the resulting r-abundance distribution, should the r-process develop in such a high-temperature environment. The temperature at which the r-process occurs remains however unknown.

\section{Conclusion}
\label{sect_concl}

The reaction code TALYS has undergone several developments. These include the modification of the code to estimate the Maxwellian-averaged reaction rates of particular relevance to astrophysics, and the inclusion of microscopic nuclear models for ground state properties, nuclear level densities, $\gamma$-ray strength functions, and nucleon-nucleus optical potentials. 
These developments enable the reaction rates to be estimated with improved accuracy and reliability, and the approximations, assumed until now by other codes, to be tested. The TALYS predictions for the thermonuclear rates  are found to be relatively similar to the more traditional astrophysics codes for stable nuclei, but major differences are obtained when considering exotic neutron-rich nuclei. In particular, the pre-equilibrium process, neglected in previous astrophysically-orientated reaction codes, is found to play an important role at low energies, at least for neutron-rich nuclei. TALYS can also estimate the multi-particle emission cross sections in photon and particle-induced reactions that have been until neglected. More specifically, the (n,2n) channel was found to become important at increasing temperatures and even to dominate, at relatively high temperatures, the (n,$\gamma$) channel, for exotic neutron-rich nuclei. The presence of such a (n,2n) barrier demonstrates that the neutron captures characterizing the r-process nucleosynthesis could not occur at temperatures above typically $T=3 \times 10^9$~K. 

TALYS presents additional advantages such as the detailed description of the decay scheme, the inclusion of accurate width fluctuation corrections, the inclusion of parity-dependent level densities or a proper description of the deformation effects. The impact of such features on reaction rates of relevance  to astrophysics and on astrophysical observables remains to be studied in the future. 

All of these developments can now be applied to improve the determination of reaction rates which  will hopefully help to place nuclear astrophysics applications on a safer footing. The TALYS predictions for the thermonuclear rates are made available to the scientific community at the {\it http://www-astro.ulb.ac.be} website  for neutron, proton, and $\alpha$-induced reactions at temperatures ranging between $10^5$ and $10^{10}$~K for all nuclei between $Z=8$ and $Z=100$ lying between the proton and neutron drip lines. The TALYS code is publicly available at {\it http://www.talys.eu}. The neutron, proton and $\alpha$-particle capture rates for all $8 \le Z \le 105$ nuclei lying between the proton and neutron drip lines are available in electronic form at the CDS via anonymous ftp to cdsarc.u-strasbg.fr or via http://cdsweb.u-strasbg.fr/cgi-bin/qcat?J/A+A/.

\begin{acknowledgement}
S.G. is F.N.R.S research assistant.
\end{acknowledgement}


\begin{thebibliography}{}
\bibitem{bruslib05} Aikawa, M., Arnould, M., Goriely, S.  et al. 2005, A\&A, 441, 1195

\bibitem{akk85} Akkerman, J.M., \& Gruppelaar, H. 1985, Phys. Lett. B, 157, 95

\bibitem[]{} Arnould, M. 1972, A\&A, 19, 92

\bibitem{bruslib06} Arnould, M., \& Goriely, S.  2006, Nucl. Phys.,  A777, 157c

\bibitem{arnould07} Arnould, M., Goriely, S., \& Takahashi, K.  2007, Phys. Rep., 450, 97

\bibitem[]{} Bao Z.Y., Beer H., K\"appeler F., et al. 2000, At. Data Nucl. Data Tables,
75, 1

\bibitem{ripl2} Belgya T., Bersillon, O., Capote Noy, R. et al. 2006, {\it Handbook for calculations of nuclear reaction data, RIPL-2} (IAEA-Tecdoc-1506)

\bibitem[]{}  Christy F., \& Duck, I. 1961, Nucl. Phys. 24, 89

\bibitem[]{} Cowan, J.J., Thielemann, F.-K., \& Truran, J.W. 1991, Phys. Rep., 208, 267

\bibitem[]{} Duijvestijn M.C., \& Koning, A.J. 2005, in Proc. of Nuclear Data
for Science and Technology (AIP Conf. Proc.  769, eds. Haight et al., p.1225)

\bibitem{go97} Goriely, S.  1997, A\&A, 325, 414

\bibitem[]{} Goriely, S. 1998, Phys. Lett., B436, 10 

\bibitem[]{} Goriely, S., \& Khan, E. 2002, Nucl. Phys., A706, 217 

\bibitem[]{} Goriely, S., Khan, E., \& Samyn, M. 2004, Nucl. Phys., A739, 331

\bibitem{go06} Goriely, S., Pearson, J.M.,  \& Samyn, M. 2007  Phys. Rev, C75, 064312

\bibitem{go08} Goriely, S., Hilaire, S.,  Koning, A.J., Sin, M. \& Capote, R. 2008, in preparation

\bibitem[]{} Hauser, W., \& Feshbach, H. 1952, Phys. Rev., 87, 366

\bibitem{hil03} Hilaire, S., Lagrange, C.,  \& Koning A.J. 2003, Ann. Phys., 306, 209

\bibitem{hil06} Hilaire, S., \&  Goriely, S. 2006, Nucl. Phys., A779, 63

\bibitem{hrtw1} Hofmann, H.M., Richert, J., Tepel, J.W.,  \& Weidenm\"{u}ller,  H.A. 1975,  Ann. Phys.,  90, 403

\bibitem{hrtw80} Hofmann, H.M., Mertelmeier, T., Herman, M., \& Tepel J.W. 1980,  Zeit. Phys., A297, 153

\bibitem[]{} Holmes, J.A., Woosley, S.E., Fowler, W.A., \& Zimmerman, B.A 1976,
Atomic Data Nucl. Data Tables, 18, 306

\bibitem[]{} Koning, A.J., Beijers, H., Benlliure J., et al., 2002, in Proc. of Nuclear Data
for Science and Technology, J. Nucl. Science and Technology, Suppl. 2, vol. 2, ed. 
K. Shibata (Atomic Energy Society of Japan) 1161

\bibitem{koning03} Koning, A.J., \& Delaroche, J.-P. 2003, Nucl. Phys., A713,  231

\bibitem{koning04} Koning, A.J., \& Duijvestijn, M.C.  2004, Nucl. Phys., A744, 15

\bibitem{koning04a} Koning, A.J., Hilaire, S., \& Duijvestijn, M. 2004, in TALYS: A nuclear reaction program (NRG-report 21297/04.62741/P) also available at {\it http//www.talys.eu}.

\bibitem{koning07} Koning, A.J., Hilaire, S., \& Duijvestijn, M. 2007, in Nuclear Data for Science and Technology, in press

\bibitem{fadden} McFadden, L., \& Satchler, G.R. 1966, Nucl. Phys., 84, 177

\bibitem[ ]{ } Mathews, G.J., Mengoni, A., Thielemann, F.-K. \& Fowler, W.A. 1983, ApJ, 270, 740

\bibitem[ ]{ } Mocelj,D., Rauscher, T., Martõnez-Pinedo  G., et al. 2007, Phys. Rev. C 75, 045805

\bibitem{moldauer80} Moldauer, P.A. 1980, Nucl. Phys., A344, 185

\bibitem[]{} Rauscher, T., Bieber, R, Oberhummer, H., 
%Kratz, K.-L., Dobaczewski, J., 
et al. 1998, Phys. Rev., 57, 2031

\bibitem[]{} Rauscher, T., \& Thielemann, F.-K. 2001, At. Data Nucl. Data
Tables, 79, 47

\bibitem[]{}  Sargood, D.G. 1982, Phys. Rep., 93, 61

\bibitem[]{} Satchler, G.R. 1983, Direct Nuclear Reactions (Clarendon press, Oxford)

\bibitem[]{}  M. Segawa, M., Masaki, T., Nagai, Y., et al. 
%Temma, Y., Shima, T., Mishima, K.,  Igashira, M., Goriely, S., Koning,  A.J., \& Hilaire, S. 
2007, Phys. Rev., C76, 022802

\bibitem{koning07} Sin, M., Capote, R., Goriely, S.,  Hilaire, S., \& Koning, A.J. 2007, in Nuclear Data for Science and Technology, in press

\bibitem[]{} Thielemann, F.-K.,  Arnould, M., \& Truran, W. 1987, in Adv. Nucl. Astrophys., eds. Vangioni-Flam E., et al. (Frontiers, Gif-sur-Yvette), p. 525

\bibitem{verbaarschot85} Verbaarschot, J.J.M., Weidenm\"{u}ller, H.A., \& Zirnbauer, M.R. 1985. Phys. Rep., 129, 367

\bibitem[]{} Woosley, S.E., Fowler, W.A., Holmes, J.A., \& Zimmerman, B.A. 1978, Atomic Data Nucl. Data Tables, 22, 371

\end{thebibliography}
\end{document}